\documentclass[journal=prE,twocolumn]{revtex4-1}
\usepackage{amsmath}
\usepackage{amsfonts}
\usepackage{graphicx}
\usepackage{epstopdf}
\newcommand{\tdot}[1]{\ensuremath{\,\begin{smallmatrix}\\({#1})\\\end{smallmatrix}\,}}

\usepackage{accents}

\begin{document}

\title{ Real-space quadrature: a convenient, efficient representation for multipole expansions}

\author{ David M. Rogers}
\affiliation{ University of South Florida, 4202 E. Fowler Ave., CHE 205, Tampa, FL 33620}
\email{ davidrogers@usf.edu}

\date{\today}

\begin{abstract}
  Multipolar expansions are a foundational tool for describing basis functions in quantum mechanics,
many-body polarization, and other distributions on the unit sphere.  Progress on these topics is often held back by complicated and competing formulas for calculating and using spherical harmonics.
We present a complete representation for supersymmetric 3D tensors
that replaces spherical harmonic basis functions by a dramatically
simpler set of weights associated to discrete points in 3D space.
This representation is shown to be space optimal.  It reduces
tensor contraction and the spherical harmonic decomposition of Poisson's operator
to pairwise summations over the point set.  Moreover,
multiplication of spherical harmonic basis functions translates to a
direct product in this representation.
\end{abstract}


\maketitle
\section{ Introduction}\label{s:intro}

 
 Multipole expansions are used throughout QM, advanced MM, and large-scale electrostatics calculations.  They appear as derivatives of $r^{-1}$ in the Green's function solution to Poisson's equation.  Applications include the computation of angular interaction matrix elements (themselves based on spherical harmonic functions) in quantum codes,\cite{gfere97,dchan08,gbelk13} fast $O(N)$ implicit solvent calculations in molecular dynamics\cite{abord03,mschn07,mschn07b}
and fluid flows,\cite{gdass02,fcruz11} describing molecular polarization\cite{stone,bjezi94,pren03,ckram13,ranan13}, and boundary representations
of dielectric polarization in molecular cavities.\cite{jappl89,ecanc97,jbard11,jbard12}

  Despite their wide-ranging applications, actual use of these expansions is hindered
by the tenuous connections between spectral (spherical harmonic) and
real-space (Cartesian) formalisms.\cite{bjezi94}
Essentially all serious applications make use of
manually tabulated Cartesian forms, often implemented by hand.\cite{jappl89,asawa89,cpozr92,gdass02,pren03,sturz11,klore12}
This strategy is error-prone, time-consuming and useful only for low-order modes.
This leads, for example, to widely-used and highly-regarded quantum chemistry
codes that only support angular basis functions up to order 6,\cite{nwchem}
and simulation methods and codes for which even implementations at the
dipole\cite{atouk00} or quadrupole\cite{pren02,dcase05} level are great advancements.
In addition, there is a gap in understanding translations from harmonics
to representations relying completely on
point-charges,\cite{ranan13} or on explicit supersymmetric tensors.\cite{pren02,ljaco09}
The straightforward route for a point-charge
representation of multipoles up to order $p-1$
requires $2^p$ charges,\cite{asawa89,dchan08}
while a tensor representation requires $3^p$ tensor elements.
In this work we show space-optimal representations for point-charges and Cartesian tensors,
and provide simple translations between these three formalisms.


  The ubiquitous success of spherical harmonics stems from the
separation of source and destination points ($x$ and $y$) in the expansion
of the potential function,
\begin{align}
\frac{1}{|y-x|} &= \sum_{l=0}^\infty \sum_{m=-l}^l O_{lm}(y) M_{lm}(x) , \quad |y| < |x| \\
\intertext{where $x$ and $y$ are three-dimensional vectors, and}
O_{lm}(x) &= |x|^l (l+|m|)! P_{lm}(\cos \theta_x) e^{-i m \phi_x} \label{e:corresp} \\
M_{lm}(x) &= |x|^{-l-1} (l-|m|)!^{-1} P_{lm}(\cos \theta_x) e^{i m \phi_x}
\end{align}
This simplified form is due to Ref.~\cite{cwhit94}.
Here, $i$, is the unit imaginary number and $P_{lm}$ are the associated Legendre polynomials.

  We contend that for almost all real applications, this decomposition goes one step too far.
Rather, it is preferable to stop at the unadorned Legendre polynomials,
\begin{equation}
P_l(\hat x \cdot \hat y) =  \sum_{m=-l}^l O_{lm}(\hat x) M_{lm}(\hat y)
,
\end{equation}
where $\hat x$ represents a three-dimensional unit vector, $|\hat x| = 1$.
The $P_l$ are polynomials of degree $l$ in the vector inner-product, $\hat x\cdot \hat y$,
and are obviously symmetric to interchanging $x$ and $y$.
No angles need to be defined or used.

  In preference to the pair, $O$ and $M$, we then use scaled Legendre
polynomials,
\begin{equation}
L_n(x,y) \equiv \frac{|x|^n}{|y|^{n+1}} P_n(\hat x \cdot \hat y) 
\label{e:harm}
.
\end{equation}
When scaled by $|y|^{2n+1}$, the $L_n$ are polynomials in the inner products,
$x\cdot x$, $y\cdot y$, and $x\cdot y$, symmetric to interchange of $x$ and $y$.
They can be generated by a two-term recurrence [since $L_n(x,y) = F_n(x,y; -1)$],
\begin{align}
 F_n(x,y;\alpha) &\equiv (-x\cdot\partial_y)^n |y|^\alpha/n! \label{e:rec} \\
 &= |y|^\alpha, \quad n=0 \notag \\
 &= -\alpha \frac{x\cdot y}{y\cdot y} |y|^\alpha, \quad n=1 \notag \\
 &= \frac{2n-2-\alpha}{n} \frac{x\cdot y}{y\cdot y} F_{n-1}(x,y;\alpha) \\
  & \;\; + \frac{\alpha+2-n}{n} \frac{x\cdot x}{y\cdot y} F_{n-2}(x,y;\alpha)
.
\end{align}
They express successive derivatives of $|y|^{-1}$, so that
\begin{align}
\frac{1}{|x-y|} = e^{-x\cdot\partial_y} |y|^{-1} = \sum_{n=0}^\infty L_n(x,y) \label{e:Pexp}
.
\end{align}

\subsection{ Definition of the Quadrature Representation}
\begin{figure}
\includegraphics[width=0.45\textwidth]{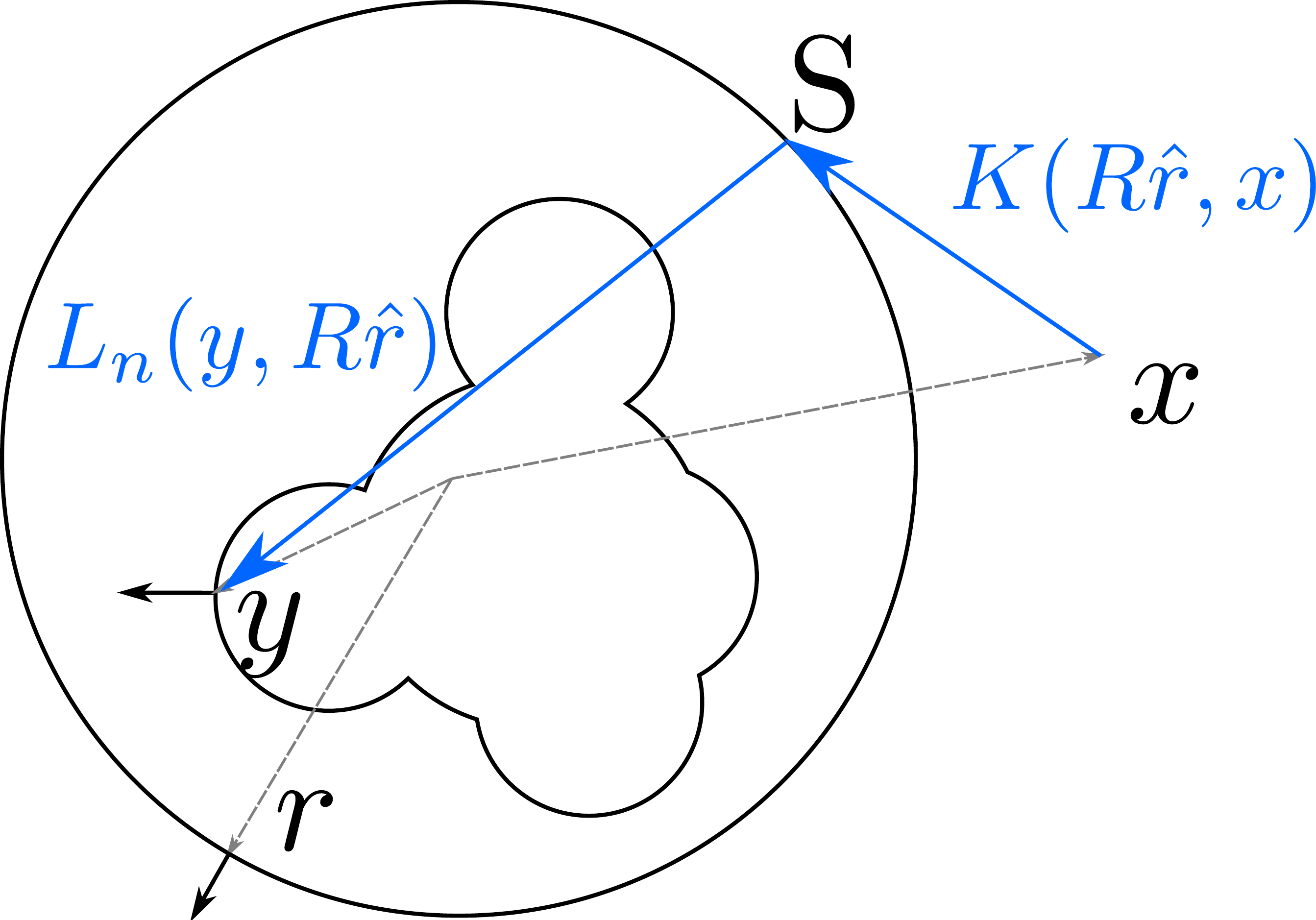}
\caption{Illustration of quadrature representation of the multipole expansion.  The source at
position $y$ is transferred to the quadrature points, $R\hat r_i$ (using Eq.~\ref{e:rep}).
Its potential inside the sphere is calculated from those effective charges (Eq.~\ref{e:Pexp}).
Normal vector directions (pointing out of the sphere) are defined for boundary
integrals.}\label{f:K}
\end{figure}

  Identities previously phrased in terms of harmonics find a much simpler expression
using the reproducing kernel (Fig.~\ref{f:K}),\cite{cahre09}
\begin{equation}
K(x,y) = \sum_{n=0}^{p-1} \frac{2n+1}{4\pi} L_n(x,y) \label{e:K}
.
\end{equation}
Its status as an effective identity kernel can be shown using the orthogonality
of the Legendre polynomials on the surface of the unit sphere ($S$),\cite{cahre09}
\begin{equation}
\int_S P_n(\hat x \cdot \hat r) P_m(\hat r \cdot \hat y) \; d^2\hat r = \delta_{nm} \frac{4\pi}{2 n + 1} P_n(\hat x\cdot \hat y)
.\label{e:ortho}
\end{equation}
As $\hat y$ is varied, the set of Legendre polynomials up to order $p-1$ span the set of all $\sum_{n=0}^{p-1} 2n+1 = p^2$ polynomials in $\hat x$ from degrees $0$ through
$p-1$ on the unit sphere.\cite{cahre09}

  Because $K$ generates polynomials, functions on $S$ can be represented
in terms of their values at a small set of quadrature points.
We make use of standard quadrature formulas prescribing
a set of $N \sim \tfrac{3}{2}p^2$ roots, $\hat r_i$, and associated weights, $w^0_i$, capable
of integrating all polynomials up to order $2p-2$.\cite{cahre09}
Using such a quadrature set, for polynomial functions $\sigma$ of degree less than $p$,
\begin{equation}
\sigma(\hat x) = \int K(\hat x, \hat y) \sigma(\hat y) \; dy
  = \sum_{i=1}^N w^0_i \sigma(\hat r_i) K(\hat x, \hat r_i)  .
\label{e:rep}
\end{equation}
We will see that the coefficients $w^0_i \sigma(\hat r_i)$ act as the effective
charges on the spherical surface.
The numerical computations presented in Sections~\ref{s:fmm} and~\ref{s:sphere} make use of
the Lebedev quadrature set.\cite{vlebe77}

  The choice of scale with $|x|$ and $|y|$ in Eq.~\ref{e:K} prescribes a default behavior off the
surface of the unit sphere coinciding with that of Ref.~\citenum{cwhit94}:
\begin{align}
\int_S K(x,\hat y) O_{lm}(\hat y) \; d^2\hat y &= O_{lm}(x) \\
\int_S M_{lm}(\hat x) K(\hat x, y) \; d^2\hat x &= M_{lm}(y)
.
\end{align}
These formulas can be used to directly translate formulas using spherical harmonics
back into Legendre polynomials, as is done in Appendix~\ref{s:harm}.
Each choice of the coordinate axes for spherical harmonics
of order $n$ generates $2n+1$ harmonics, which correspond to
a particular basis for the $2n+1$ degenerate eigenfunctions of
$P_n(r_i\cdot r_j)$.  According to Eq.~\ref{e:ortho}, each of the
$2n+1$ vectors has eigenvalue $4\pi/(2n+1)$.  Working directly
with $P_n$ avoids decomposition to this basis, removing angular
variables from the resulting expressions.

  The required separation of source and destination points ($x$ and $y$
in Fig.~\ref{f:K}) is easily achieved using the scaled reproducing kernel,
\begin{align}
(-y&\cdot\partial_x)^n |x|^{-1}/n! = L_n(y,x) \notag \\
  &= \int_S K(y/R, \hat r) L_n(R \hat r, x) \; d^2\hat r \label{e:outer2} \\
\intertext{or}
  &= \int_S L_n(y, R \hat r) K(\hat r, x/R) \; d^2\hat r \label{e:inner2}
 .
\end{align}

  Equation~\ref{e:outer2} generates the {\em outer} expansion by collecting the
integration over sources inside the bounding sphere ($|y| < R$) into an effective
surface charge distribution,
\begin{align}
\sigma_o(R \hat r) &\equiv \int \rho(y) K(y/R, \hat r) \; d^3y
. \label{e:outer} \\
\intertext{The source's $n$-th order contribution to the potential at point $x$ is then}
\int L_n(y,x) \rho(y) \; d^3y &= \int_S L_n(R\hat r, x) \sigma_o(R\hat r) \; d^2\hat r
. \label{e:Gouter} \\
\intertext{Likewise, Eq.~\ref{e:inner2} generates the {\em inner} expansion when collecting
the integration over sources outside the bounding sphere, $|x| > R$,}
\sigma_i(R \hat r) &\equiv  \int \rho(x) K(\hat r, x/R) \; d^3y
.  \label{e:inner} \\
\intertext{with the $n$-th order potential inside the sphere given by}
\int L_n(x,y) \rho(x) \; d^3x &= \int_S L_n(x, R \hat r) \sigma_i(R \hat r) \; d^2\hat r
. \label{e:Ginner}
\end{align}

\subsection{ Energy in the Quadrature Representation}

  The interaction energy between two sets of point sources ($y$ inside and $x$
outside of a sphere of radius $R$) is written in a multipole expansion as
\begin{equation}
E = \sum_{x,y} \frac{q_x q_y}{|x - y|} = \frac{1}{R} \sum_{x,y,n=0} q_x q_y L_n(y/R,x/R)
\end{equation}
Using both the outer expansion projecting the enclosed charges
onto a surrounding sphere ($L_n(y,x) \to \int_S K(y/R,\hat r) L_n(R\hat r, x) \; d^2\hat r$, Eq.~\ref{e:outer2}),
and the inner expansion projecting the outer charges onto the same sphere
($L_n(R\hat r, x) \to \int L_n(R\hat r, R\hat s) K(\hat s, x/R) d^2\hat s$, Eq.~\ref{e:inner2}),
\begin{align}
E &= \frac{1}{R} \sum_{x,y,i,j,n=0} L_n(\hat r_i, \hat r_j) w^0_i w^0_j \notag \\
 & \qquad \times q_y K(y/R,\hat r_i) q_x K(\hat r_j, x/R) \notag \\
 &= \frac{1}{R} \sum_{i,j,n=0} w^0_i \sigma_o(R\hat r_i) w^0_j \sigma_i(R\hat r_j) L_n(\hat r_i, \hat r_j)
 .
\end{align}
The second step uses the definitions in Eq.~\ref{e:outer} and Eq~\ref{e:inner}.
In most cases, expressions involving the potential are simpler.

  In addition, when $R$ is large, the energy is described equally well
by treating the spherical quadrature points as point charge sources.  This can be derived by
inserting the moment shifting formula (Eq;~\ref{e:ishift}) for
$\sigma_i(R \hat r_j) \to \sum_k K(\hat r_j, \frac{R_x \hat r_k - t}{R}) w^0_k \sigma_o(R\hat r_k;t)$
and summing over $j$ to show
\begin{align}
E &= \sum_{i,k,n} w^0_i \sigma_o(R\hat r_i) w^0_k\sigma_o(R_x \hat r_k; t) L_n(R\hat r_i, R_x \hat r_k - t) \\
  &\simeq \sum_{i,k} \frac{w^0_i \sigma_o(R\hat r_i)w^0_k \sigma_o(R_x \hat r_k; t) }{|R_x r_k -t -R \hat r_i|}
\end{align}

\subsection{ Outline}

  Much of the work focused on real-space (Cartesian) representations of the
expansion (Eq.~\ref{e:Pexp}) uses a tensor notation.  The set of three-dimensional tensors of orders n, $\sum_i q_i r_i^{(n)}$, from $n=0$ up to order $n=p-1$ is termed a `Cartesian polytensor.'
Appendix~\ref{s:tens} shows a direct correspondence between Cartesian polytensors
and polynomial functions with maximum degree $p-1$.
It is shown that outer and inner products between supersymmetric tensors
have much simpler statements in terms of polynomial multiplication and differentiation.
Moreover, the set of trace-free Cartesian polytensors\cite{jappl89,pren03} is
exactly identified with the set of polynomials on the unit sphere, $\sigma(\hat x)$.
In combination with the representation
theorem, Eq.~\ref{e:rep}, these identities completely connect tensor and
harmonic representations to a novel, quadrature representation of polytensor space.

  Section~\ref{s:fmm} illustrates the simplicity of the quadrature method by testing
moment shift formulas used during the fast multipole method.
Surprisingly, these have the same form as the initial moment fitting.
Moreover, it shows that the weights for the quadrature representations exactly
coincide with point charges that reproduce those multipole moments
at sufficient distance from the bounding sphere.
As expected by the exact correspondence to
spherical harmonics, numerical results show that the
error of representing random point sources scales with distance
identically to traditional spherical harmonic multipoles.
Section~\ref{s:sphere} provides exact quadratures for the single and
double-layer potentials on a spherical surface.  Numerical results in
Sec.~\ref{s:sphere} demonstrate a singularity representation for multiple spheres
interacting through a perfect, irrotational fluid.

  The conclusion summarizes the connections created here
and outlines new applications and simplifications that can be
tackled in future work.


\section{ Conversion between Quadrature and Cartesian Tensors}

  Although the quadrature representation is a complete, self-contained
basis for expressing multipole moments and externally imposed fields,
comparison with existing literature requires translation between Cartesian
representations.

  Writing the order-$n$ Cartesian multipole moment tensor as
$M^{(n)} \equiv \int y^{(n)} \rho(y)\; d^3y$,
the directional moment is defined as the complete ($n$-way) tensor contraction
between $M^{(n)}$ and a test vector, $r$,
\begin{equation*}
r^{(n)} \tdot{n} M^{(n)} = \int (r\cdot y)^n \rho(y) \; d^3y
.
\end{equation*}

  Complete information about the polytensor, $\{M^{(n)}, n=0, \ldots, p-1\}$ is
contained in the set of directional moments against the quadrature points, $\{\hat r_i\}$
(see Appendix~\ref{s:tens} for details).  These directional moments
can be substituted wherever powers of ($\hat r\cdot y$) appear
in the moment matching equation~\ref{e:outer}.

  Conversion back to Cartesian form is also simple, since the quadrature
weights were defined to reproduce polynomial integrals, which include
all (trace-free) tensors,
\begin{equation}
\int \mathcal D_n r^{(n)} \rho(r) \; d^3r = \sum_i w^0_i \sigma(\hat r_i) r_i^{(n)},\, n < p
.
\end{equation}
Here, $\mathcal D_n$ is the de-tracer projection of Ref.~\citenum{jappl89}
defined by
\begin{equation}
\hat x^{(n)} \tdot{n} \mathcal D_n \hat y^{(n)} = \frac{n!}{(2n-1)!!} P_n(\hat x\cdot \hat y)
\label{e:D}
\end{equation}

  These formulas show that the conversion to quadrature form is one way to
project a polynomial into a trace-free form.
Alternatively, trace-free Cartesian tensors simplify Eq.~\ref{e:outer}, since
Eq.~\ref{e:D} can be used in $K$, (Eq.~\ref{e:K})
\begin{align}
\sigma_o(R\hat r) = \sum_n &\frac{(2n+1)}{4\pi}\frac{(2n-1)!!}{n!} \notag \\
  &\times \hat r^{(n)} \tdot{n} \mathcal D_n M^{(n)} R^{-n}
  .
\end{align}
This sum includes only the leading terms of $L_n$ at each order, $n$.

\section{ FMM Operations}\label{s:fmm}

  In this section, we elaborate the translation formulas for the quadrature representation
of multipoles.  We achieve greater generality in comparison with
previous Cartesian formulations\cite{jmaki99},
by working directly with the moment space, rather than the potential.  In comparison
with black-box fast multipole methods,\cite{lying04} the availability of spherical quadrature rules justifies the use of Legendre polynomials because they represent moment space --
regardless of their specific role in the Poisson problem.  For a detailed introduction
to the fast multipole method, see Ref.~\cite{yliu06}.

  The outer expansion is defined by the moments of the enclosed charge
distribution.  The moments of the shifted distribution should coincide with those of the original.
This can be shown from
\begin{equation}
\sigma_o(R_1\hat r; x_1) = \int_S K\Big(\frac{R_0\hat s + x_0-x_1}{R_1}, \hat r\Big) \sigma_o(R_0\hat s; x_0) \; d^2\hat s
.\label{e:shift}
\end{equation}
The notation here is that $\sigma_o(R_1\hat r; x_1)$ represents the outer expansion
about the origin $x_1$, defined on the set of points at a distance $R_1$ from $x_1$.
Since both outer expansions are polynomials of maximum degree $p-1$ in $R \hat r$,
the reproducing kernel, $K$ is able to duplicate them from any appropriate set
of quadrature points.

  The inner expansion is defined by the derivatives of the potential about the origin.
This makes translation troublesome because the expansion about a new point
does not necessarily need to obey the trace-free condition.
An implicit argument can be used to show that propagating the inner expansion to the potential
on a small sphere around the new expansion point can be used to find charges on
a new enclosing sphere that duplicate this potential up to the same order
as the original expansion.  The net result for the inner expansion is the same
as the initial fitting,
\begin{align}
\sigma_i(R_1 \hat r; x_0+t)
  &= \int K(\hat r, \tfrac{R_0\hat s - t}{R_1}) \sigma_i(R_0 \hat s; x_0) \; d^2\hat s
  \intertext{or}
    &= \int K(\hat r, \tfrac{R_0\hat s - t}{R_1}) \sigma_o(R_0 \hat s; x_0) \; d^2\hat s \label{e:ishift}
 .
\end{align}
Both outer $\to$ inner and inner $\to$ inner shift operators give the same equation.
In fact, all shift operators arriving at an outer expansion are identical to
the initial moment matching (Eq.~\ref{e:outer}).  All shift operators arriving at an inner
expansion are also identical to the inverse moment matching (Eq.~\ref{e:inner}).
Appendix~\ref{s:harm} shows that these formulas are identical to
traditional expressions involving spherical harmonics.

  The only caveat that appears here is that the kernel, $K(\hat r, \hat s - t)$, contains
singularities when $|t| = 1$.  This problem enters because the
shifting formula is only well-defined when $t$ 
does not approach the sphere bounding the source charge distribution.
This motivates the introduction of the scaled distributions in Eq.~\ref{e:outer}.  The
use of unscaled moments in shift formulas based on spherical-harmonics,
while convergent for finite sums, might therefore be expected to show numerical
issues in the limit of large expansion order, $p$.\cite{jmaki99,cless12}

\begin{figure}
\includegraphics[width=0.45\textwidth]{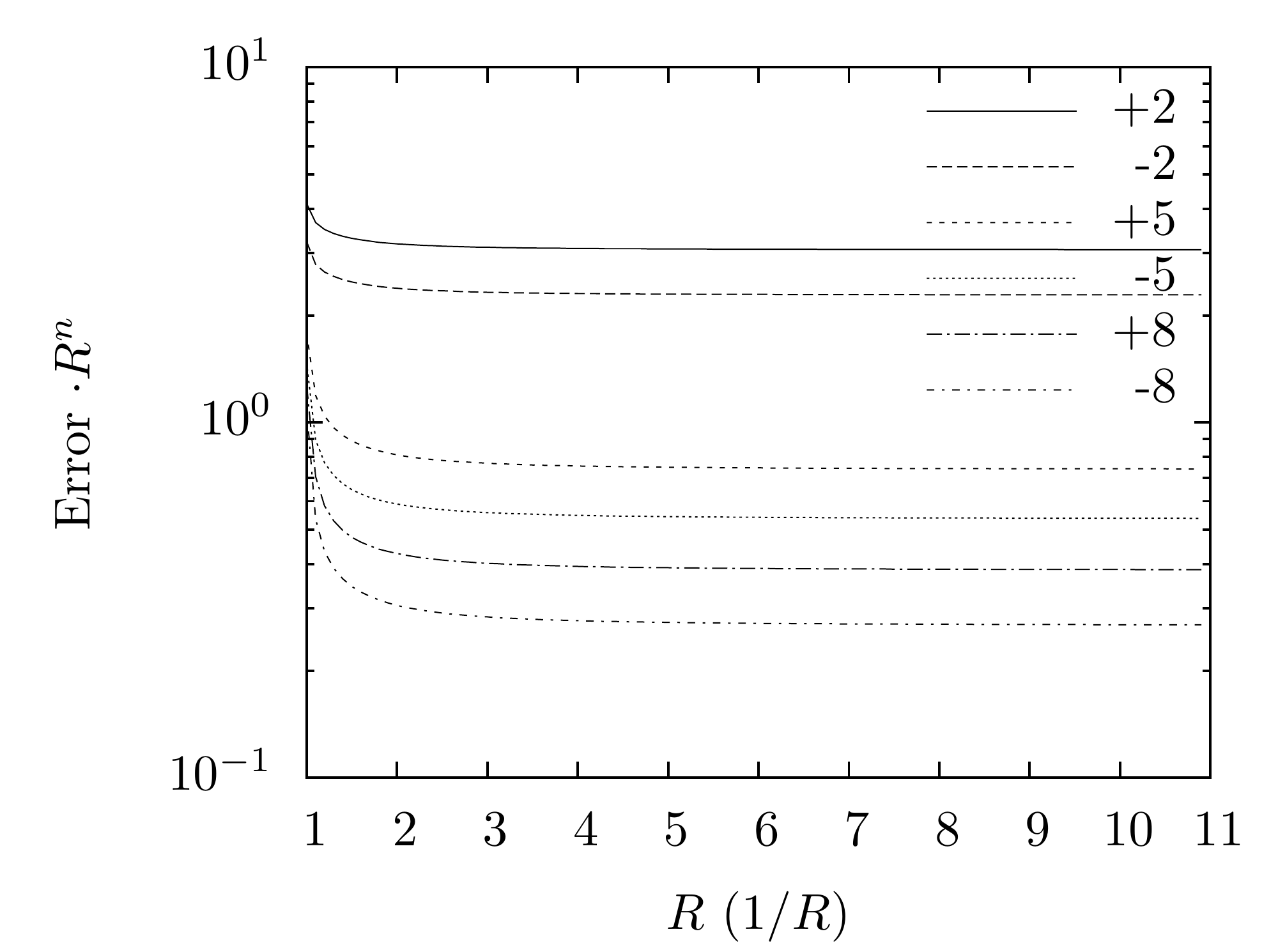}
\includegraphics[width=0.45\textwidth]{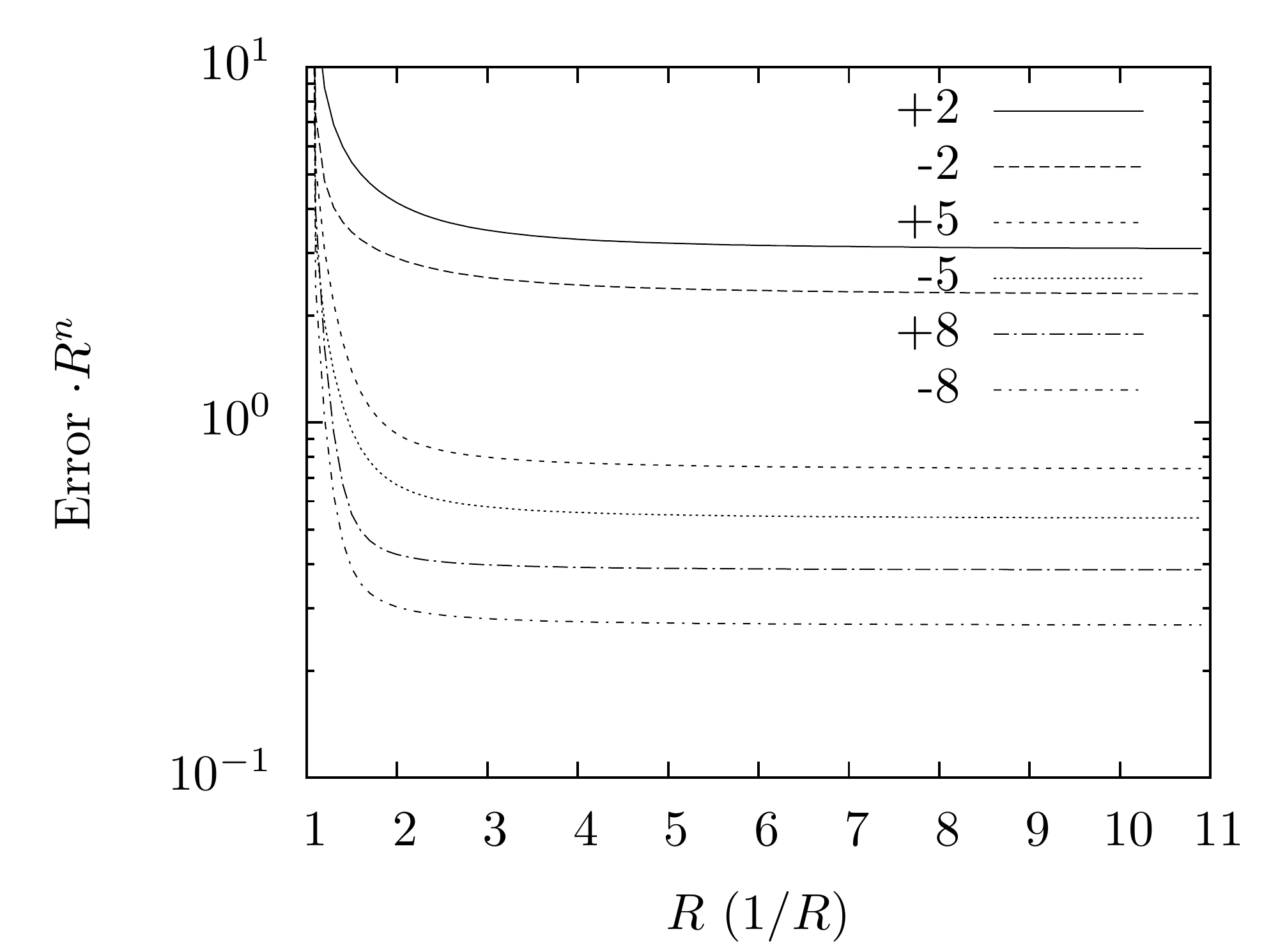}
\caption{Pre-factors for the multipole representation error.
The residual is scaled by $r^{p+1}$ (outer expansion, labeled $+p$) or $r^{-p}$ (inner expansion, labeled $-p$).
The left panel shows the error of exact quadrature using Eq.~\ref{e:Gext} or Eq.~\ref{e:Gint},
while the right shows the error of summing $1/r$, treating the quadrature points as point sources.  Point charge sources make excellent representations of multipole sources and external fields.}\label{f:racc}
\end{figure}

    Figure~\ref{f:racc} shows that the numerical error of the quadrature representation, including
appropriate bounding radii, have the same power-law decrease in error
as previously shown for spherical harmonics.
It plots the accuracy of representing the electrostatic potential
as a function of distance from the bounding sphere.
A set of 4000 point charges assigned from a uniform distribution
between -1 and +1 were placed uniformily in the cube $[-\frac{\sqrt 3}{3},\frac{\sqrt 3}{3}]^{(3)}$.
The residual error per point shown was averaged over $100$ charge distributions
and over the 86 points of the Lebedev quadrature grid of order 15 on the surface
of the evaluation sphere at varying distance, $r$, from the bounding sphere.
The inner expansion has an inverted geometry, with
the sources scaled by $1/|x|^2$ to put them outside the sphere,
and the evaluation test points likewise inverted to the inside, shrinking toward $r=0$.

  For comparison, the simple summation of $\sum_i w_i/ (r-R\hat r_i)$ using the quadrature
weights as charges is shown in the right panel of Fig.~\ref{f:racc}.
The pre-factor for both error curves converges at a radius above 3, showing that
the quadrature-based representation gives highly accurate point charges
that simultaneously represent all multipoles.  Practically, this solves the problem
of placing $O(p^2)$ discrete charges to mimic all multipolar moments up to
arbitrary order posed in Refs.~\cite{asawa89,gfere97,ranan13}.
This is also the reason for the symmetry of the shifting formulas -- the
weights are arrived at through the same process of fitting the inner or outer expansion.

\begin{figure*}
\includegraphics[width=0.8\textwidth]{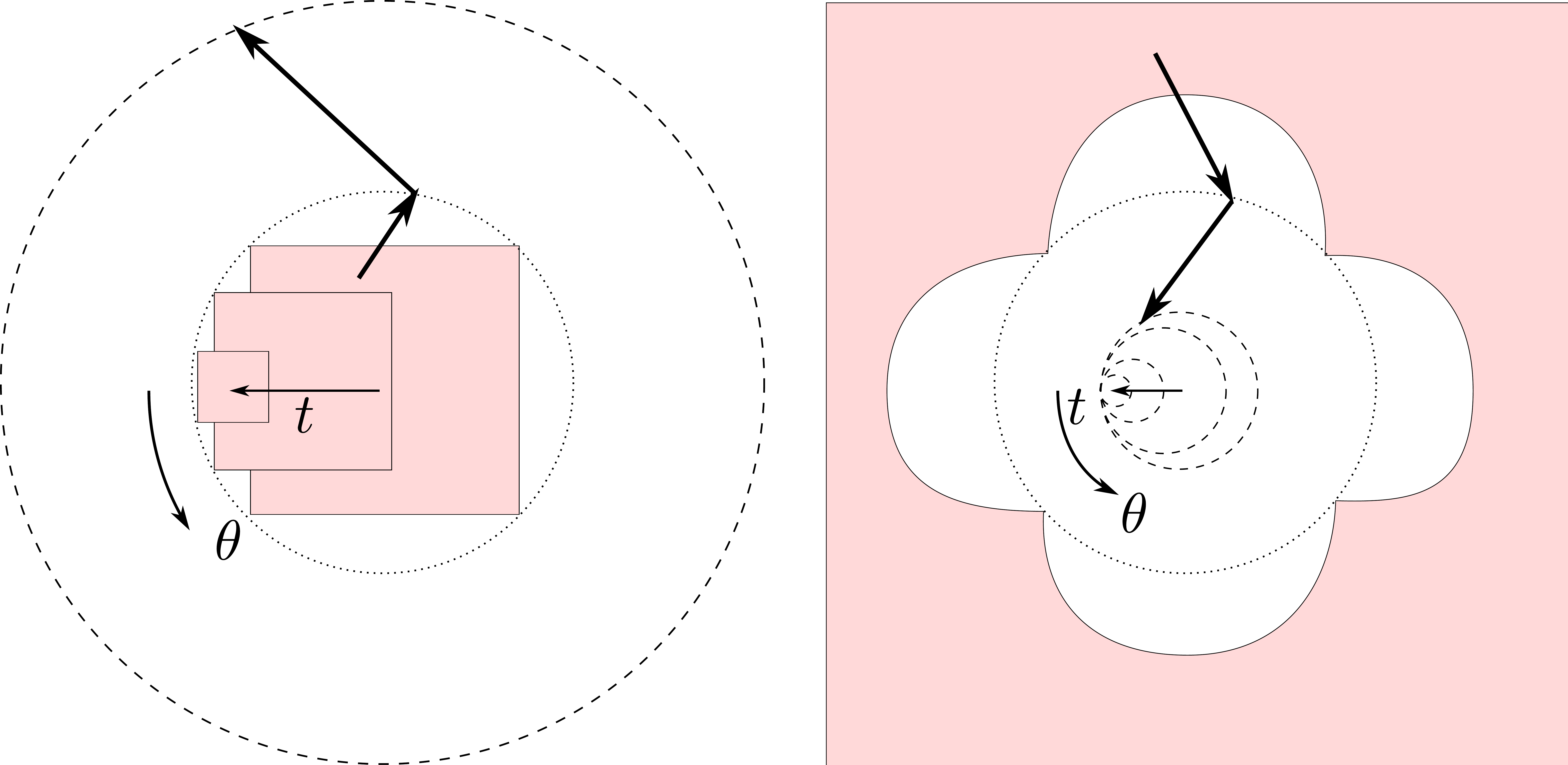}
\includegraphics[width=0.45\textwidth]{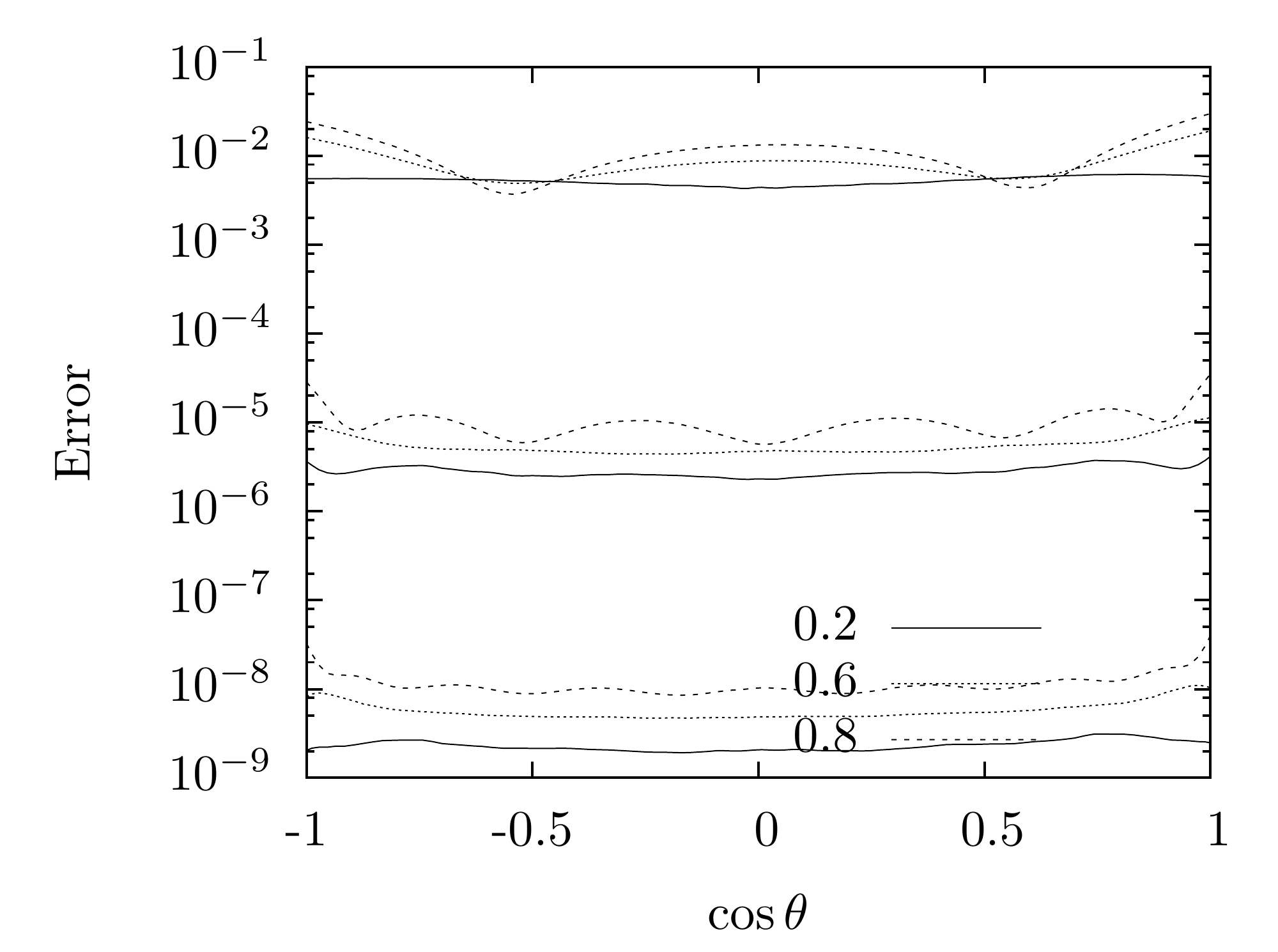}
\includegraphics[width=0.45\textwidth]{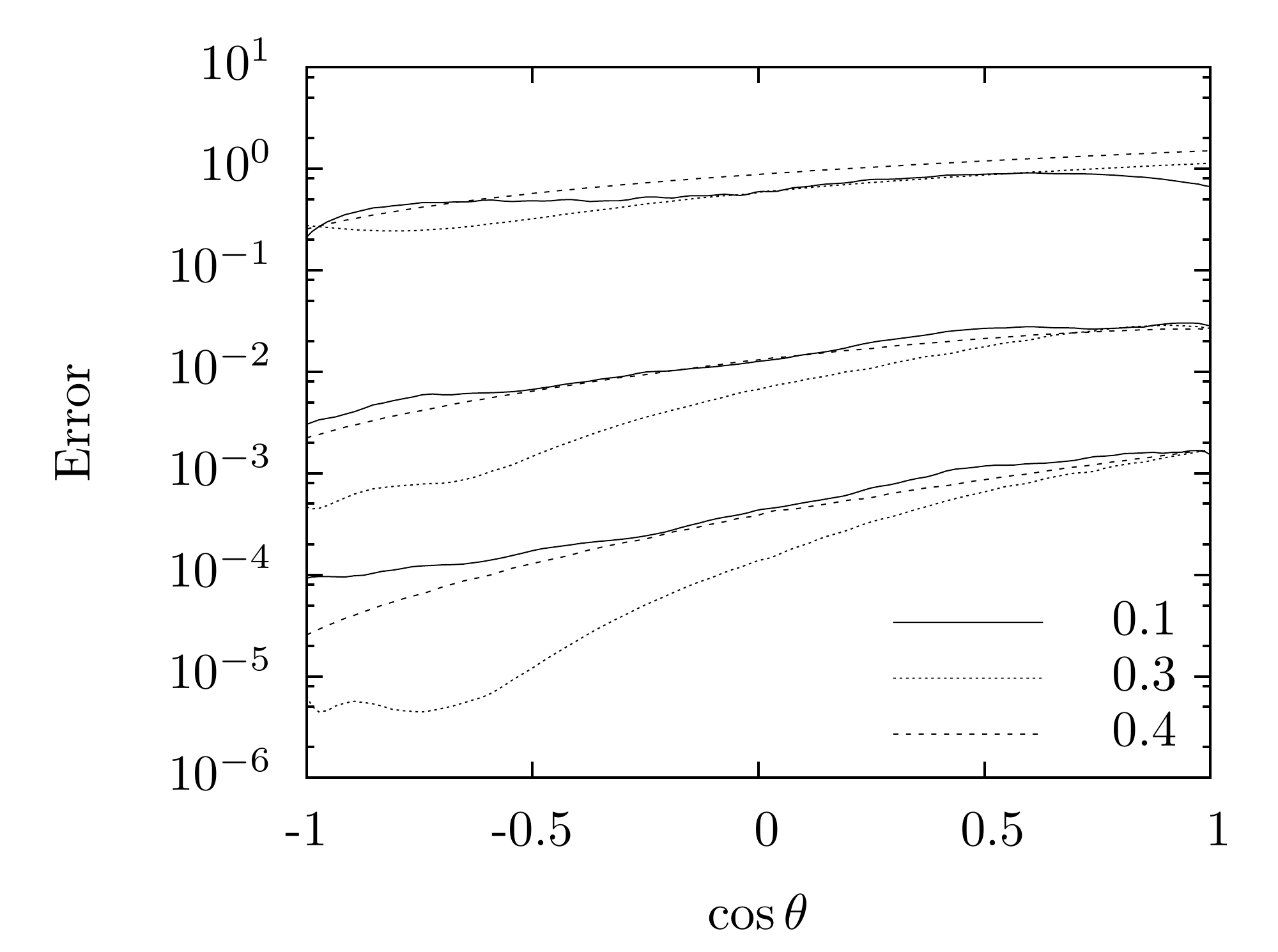}
\caption{Absolute multipole representation error in
the potential for shifted source distributions in the outer expansion ($|x|=2$, left panel)
or shifted evaluation spheres in the inner expansion ($|y|=1/2$, right panel).
Graphics above each indicate the source locations, distributed randomly
throughout the shaded areas,
and evaluation points using dashed lines.  The dotted circle indicates the
bounding sphere at $R=1$.
Both are plotted as a function of $\cos\theta = \hat r\cdot \hat t$, where $\hat t$ is the shift direction.
The magnitude of the shifts are shown in the figure legend.
Expansion orders are 2, 5, and 8 as in Fig.~\ref{f:racc}.  Increasing expansion orders
show up as groups of curves with decreasing error.}\label{f:tacc}
\end{figure*}

  Figure~\ref{f:tacc} shows the error in multipole shifting operations, plotted as a function
of angle from the shift direction.  For the outer expansion, the test points
are fixed at $R=2$, while the source distribution is shifted to the right by 0.2, 0.6, and 0.8.
The locations of the source points were scaled down for each shift to maintain contact of the rightmost
face of the source cube with the unit bounding sphere.
Summation of $\sum_i w_i/(r-\hat r_i)$ from the representation weights had nearly identical error (not shown).
For the inner expansion, the source points were inverted to lie outside the sphere and remain fixed.
The evaluation sphere started at the origin with $R=1/2$, and was successively
shifted in the $x$-direction by 0.1, 0.2, 0.3, and 0.4, with the radius scaled down to maintain
contact between the rightmost point of original evaluation sphere and the shifted version.

  The error of the inner expansion shows much more variation as a function of the cosine with respect
to the shift direction, since the distance to the unit sphere is more quickly varying than in
the outer expansion geometry.  The scaling of the outer expansion with expansion order, $p$,
also appears somewhat better because the source locations were scaled down
with increasingly large shifts in that geometry.

\section{ Singularity Methods}\label{s:sphere}

  The boundary integral method utilizes the fact that in regions, $\Omega$,
where $\mathcal L \phi = 0$,
the potential can be expressed as a boundary integral using the divergence theorem,
\begin{align}
\phi(x) = \int_{\partial\Omega} \phi(y) &n\cdot p(y)\partial_{y} G(x-y) \notag \\
  &- G(x-y) n\cdot p(y) \partial_{y} \phi(y) d^2y
. \label{e:BEM}
\end{align}
Here $n$ is the inward-pointing normal vector on the surface at point $y$, and a
Sturm-Liouville form has been
assumed for $\mathcal L G \equiv \partial\cdot(p \partial G) - u G = \delta(r)$.  The normal
vector directions are shown in Fig.~\ref{f:K}.
This equality makes it possible to solve $\mathcal L \phi = 0$
with constant potential or constant flux boundary conditions.\cite{cpozr92}

  Boundary integrals over polynomial distributions on the sphere, $\sigma$, can
be computed exactly by the summations,
\begin{widetext}
\begin{align}
\int_{S} |R\hat r-x|^{-1} \sigma(R\hat r) \; d^2\hat r &= \sum_{m=0,i}^{p-1} L_m(R\hat r_i, x) w_i^\sigma \label{e:Gext} \\
\int_{S} \left[ \hat r\cdot\partial_r |R\hat r-x|^{-1}\right] \sigma(R\hat r) \; d^2\hat r &= \sum_{m=0,i}^{p-1} \frac{m}{R} L_m(R\hat r_i, x) w_i^\sigma
\label{e:Fext} \\
\int_{S} |R\hat r-y|^{-1} \sigma(R\hat r) \; d^2\hat r &= \sum_{m=0,i}^{p-1} L_m(y, R \hat r_i) w_i^\sigma \label{e:Gint} \\
\int_{S} \left[ \hat r\cdot\partial_r |R\hat r-y|^{-1} \right] \sigma(\hat r) \; d^2\hat r &= -\sum_{m=0,i}^{p-1} \frac{m+1}{R} L_m(y, R\hat r_i) w_i^\sigma
\label{e:Fint}
\end{align}
\end{widetext}
For $|y| < R$ and $|x| > R$, and where $w_i^\sigma \equiv w^0_i \sigma(R \hat r_i)$.
These can be proven by expanding $|r-x|^{-1}$ in orthogonal polynomial spaces, $L_n$, to give an integral over $S$
and then writing that integral as a quadrature, exact for polynomial $\sigma$ of degree less than $p$.
Equations~\ref{e:Gext} and~\ref{e:Gint} are just the inner and outer expansions.

  Equations~\ref{e:Fext} and~\ref{e:Fint} give the complete, singular surface integral,
not just the Cauchy principal value part.\cite{cpozr92}
This can be seen by noting that the jump discontinuity in $\hat r\cdot\partial \phi$ as
both $x$ and $y$ approach the surface point, $y$ is
\begin{align}
&F_\text{ext}\big |_{x=y} - F_\text{int} \\
&= R^2 \int_S \hat r\cdot\partial_r \frac{1}{|R\hat r-x|} \sigma(R\hat r) - \hat r\cdot\partial_r \frac{1}{|\hat r-y|} \sigma(R \hat r) \; d^2\hat r \notag \\
 &= 4\pi R \sum_n \frac{2n+1}{4\pi} L_n(R \hat y, R \hat r_i) w^0_i \sigma(R\hat r_i) = 4\pi \sigma(y)
.
\end{align}
The last equality comes from recognizing the expression for the reproducing kernel (Eq.~\ref{e:K}),
which scales as $R^n$ in its first argument and $R^{-n-1}$ in its second.

  For a set of spheres moving through an incompressible, irrotational fluid,
the flow at every point in the fluid can be written as the derivative of a potential,
\begin{align}
v(x) &\equiv -\partial \Phi(x) \label{e:vel} \\
\partial^2 \Phi(x) &= 0 \notag
.
\end{align}
Each point on a spherical boundary gives a constant field condition,
\begin{equation}
n\cdot v_0 = -n\cdot\partial \Phi(x). \label{e:bound}
\end{equation}  
The solution can be found numerically using a point-based, quadrature representation of $\Phi$,
so that Eq.~\ref{e:BEM} becomes the linear equation,
$(I-F)\cdot\Phi = G\cdot (n\cdot v_0)$.

\begin{figure}
\includegraphics[width=0.45\textwidth]{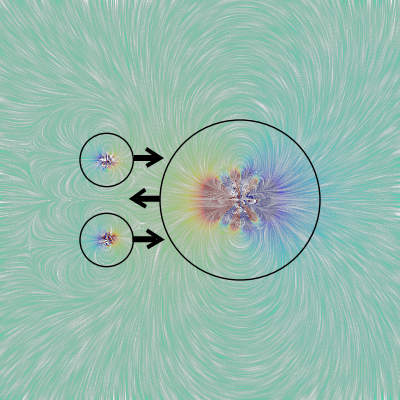} \\
\includegraphics[width=0.45\textwidth]{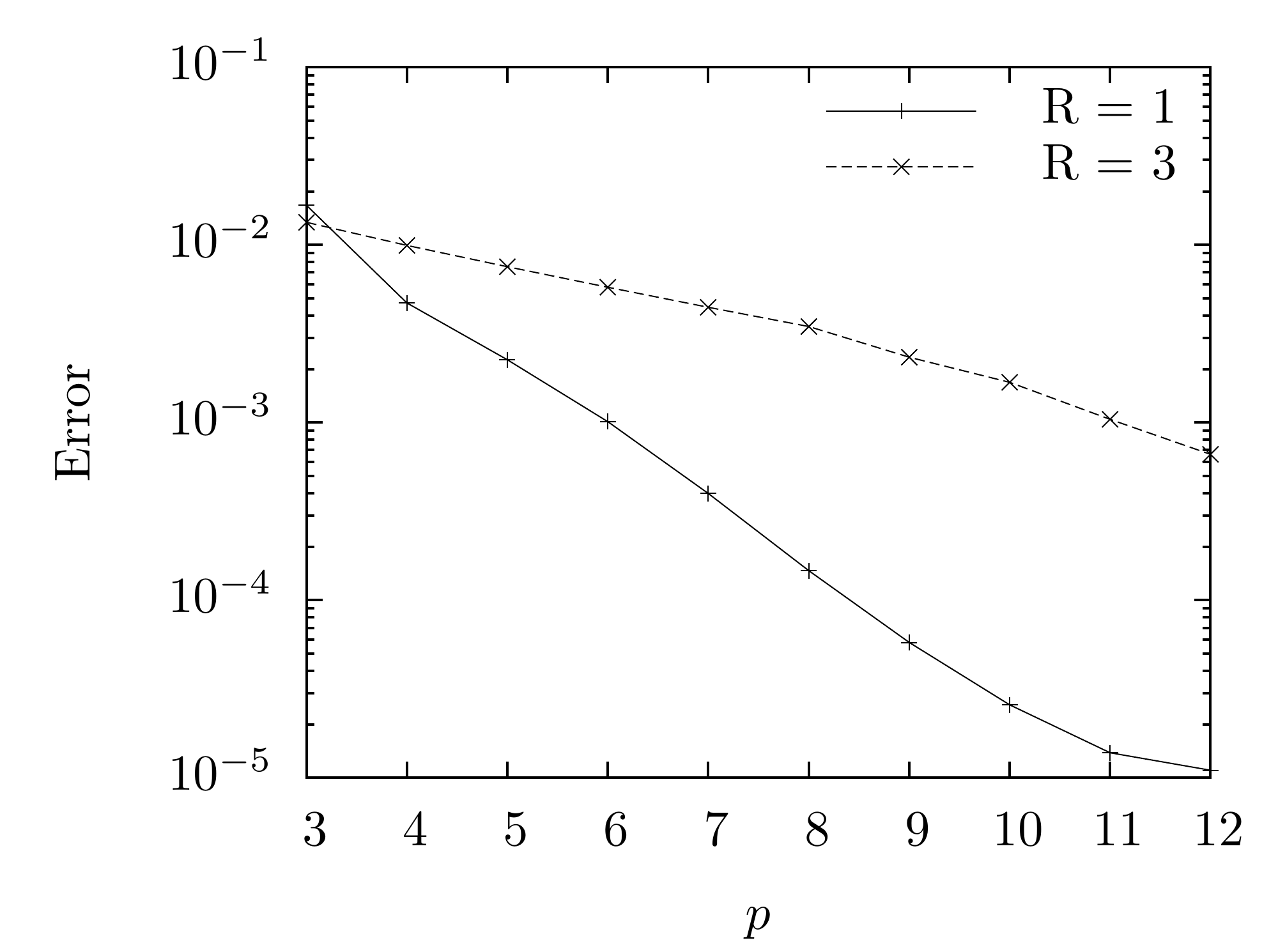}
\caption{Multipole representation of potential flow for three spheres in the $z=0$ plane (upper panel, $p=8$).  Color indicates the potential field.  The flow velocity was rendered using line integral convolution.
The lower panel shows the surface-averaged boundary error divided into contributions
from the larger and smaller spheres.}\label{f:sacc}
\end{figure}

  Figure~\ref{f:sacc}'s top panel shows the motion of three spheres through an
incompressible 3D fluid.
Point singularities represented using quadrature multipole expansions on the surfaces of
the spheres show exponential convergence.
The smaller spheres have radius 1, are located at $x=-1$ and $y=\pm 3/2$ and are traveling with unit velocity in the $+x$ direction.  The larger sphere has radius 3, is located on the $y$ axis at $x=4$,
and is traveling with unit velocity in the $-x$ direction.
The velocity at every point in the fluid is calculated using Eq.~\ref{e:vel}
from a sum of the three multipoles (outer expansion of Eqns.~\ref{e:Gext}
and~\ref{e:Fext}).  Weights on the quadrature set over each sphere were solved
to fix the boundary condition, Eq.~\ref{e:bound} at each point in the quadrature set.

  Error in the normal derivative boundary condition (lower panel of Figure~\ref{f:sacc})
was evaluated at the spherical surfaces
using a larger Lebedev quadrature grid of order 59 with 1202 points.
As the quadrature order ($p$) is increased,
more points are added to the multipolar expansion on each sphere,
and the discrepancy between the
computed and imposed normal velocity decreases exponentially.


\section{ Discussion}

  An isomorphism between symmetric Cartesian tensors and polynomial functions is used to show an optimal representation
for polytensors in terms of real weights, $w_i$, on a fixed set of basis vectors, $\{r_i\}$.  This equivalence
is compactly expressed in terms of equivalent representations of the moment integrals,
\begin{align*}
r^{(n)}\tdot{n} \int x^{(n)} \rho(x) \; d^3x &= \int (r\cdot x)^n \rho(x) \; d^3x \\
  & = \sum_i^{N} (r\cdot r_i)^n w_i
.
\end{align*}

  Three numerical results were presented.  First, we showed the numerical error in matching
spherical moments of a cloud of discrete point charges.
We compared the exact Eq.~\ref{e:Gext} with the numerical summation over the scaled
quadrature points, $R\hat r_i$.  These made use of Lebedev quadrature rules\cite{vlebe77} tabulated
up to order 131 by Burkardt.\cite{jburk}
Next, we showed that the expressions for translating the origin of
inner and outer moment expansions have the same error as the corresponding
translation formulas using spherical harmonics.  Appendix~\ref{s:harm}
gives further details on the mathematical translation.
Finally, we presented convergence results for FMM solutions to potential
flow between multiple interacting spheres.
The central role of the quadrature representation
in all these methods allows us to skip over consideration of an intermediate
spherical harmonic representation.

  When used in a Taylor expansion of a scale-invariant Green's function, $G(c r) = c^{\alpha} G(r)$,
the $r$-dependence of the solutions are fixed, and only a subset of the full moment space is required.
That subspace is equivalent to the set of polynomials
on the unit sphere.  This condition reduces the $\binom{n+2}{n}$ polynomials at
each total degree $n$ to only $2n+1$.
Approximation of integrals involving $G$ up to order $p-1$ thus requires only $N=p^2$ coefficients.
Convenient expressions for finding these coefficients from a source distribution, translation operators, and integrals involving expansions of the Poisson kernel, $|r|^{-1}$ were given.
Numerical results verified that the outer expansions have error $r^{-p-1}$,
and the inner expansions have error $r^p$,
consistent with the corresponding spherical harmonic formulas.

    The computational scaling of na\"{\i}vely evaluating the moment
translation formula Eq.~\ref{e:shift} is $p^4$.  This cost
scales as $p^3$ for spherical harmonic translations that perform rotations to align
the translation axis with the azimuthal reference axis, $\hat z$.\cite{ahaig11}.
Similar savings can be realized using the quadrature-based scheme,
when using equally spaced points
along rings spaced vertically according to Gauss-Legendre\cite{edarv00}
or Gauss-Jacobi\cite{mreut09} quadrature.
Fast, $O(p^2 \log p)$, transformations between harmonics and the Gauss-Jacobi
quadrature representation are available.\cite{mtyge08,mreut09}
Further, rotations that make use of these transformations can
achieve $O(p^3)$ scaling or, using approximate algorithms for
matrix sparsification, $O(p^2 \log p)$.\cite{edarv00}
Since the matrix in Eq.~\ref{e:shift}, $[K (\frac{R_0\hat r_j + t}{R_1}, \hat r_i )]_{ij}$
has as many unique elements as unique cosines, $\hat r_i\cdot \hat r_j$,
quadrature rules like those above containing
$O(p)$ vertical rings have a translation scaling as $O(p^2)$
after rotation.

  The Cartesian representations found in this work generalize and extend analogous results
found using manually tabulated spherical harmonics.  The decomposition of Legendre polynomials
into spherical harmonics (Eq.~\ref{e:harm}) shows that the harmonics correspond to
particular choices for the eigenfunctions of $P_n$.  To work with this set,
the matrices $[P_n(\hat r_i, \hat r_j)\sqrt{w^0_iw^0_j}]$ can be numerically diagonalized.
This gives rise to $2n+1$ eigenvectors for each $P_n$, with the same eigenvalue,
$4\pi / (2n+1)$.
Using the path from tensors to quadrature points to harmonics,
results found using Cartesian polytensors can be translated to
spherical harmonics and vice-versa.
Moreover, Cartesian tensor contractions which were na\"{\i}vely $O(3^p)$ have
been reduced to an efficient summation over an $O(p^2)$ point set.

  For differential equations not respecting scale-invariance, the representation of $\rho$
via the polynomial form of its moments still provides a simple path for deriving accurate numerical methods.
Using any $\sum_n \binom{n+2}{n}$ points for which the space of directional moments,
$[c_K(r_i; r_j)]$, has full
rank, a reproducing kernel and polynomial interpolation formulas can be found through numerical inversion
of $[c_K(r_i; r_j)]$.\cite{cdebo92}  Of course, considering only the moments actually appearing in the expansion of $G$,
using an optimal quadrature rule, and accelerating the evaluation of the moments by exploiting symmetry
will reduce the computational overhead of this exercise.

  Real-space expressions are the most desirable starting point for deriving
fast algorithms.  First, all parts of the computation have a regular, vector structure,
making simplifying optimizations for parallel hardware.
Second, high-level form of the matrix operations and lack of angular coordinates
makes derivations much simpler.
Not only are the full set of (already well-developed and efficient) spectral methods available for fast
implementation, but also new symmetry-based, sparse factorizations remain to be explored.\cite{segne01}
Sparse factorization is at the heart of fast Fourier transformation methods.
Finally, stable, $O(p^3)$ rotation formulas using point-based evaluation of the reproducing kernel (Eq.~\ref{e:K})
have been shown and critically tested in Ref.~\cite{cless12}.  Even still, a large fraction of computation
time in that work was spent on evaluating spherical harmonic functions on the set of rotated real-space points.
Either of these optimizations would reduce the overall scaling to $O(p^3)$,
recovering the scaling of rotation-based translations for spherical harmonics.\cite{lgree97}

\section{ Conclusions}

  The representations found in this work greatly simplify the analysis and numerical use
of distribution functions on a sphere.
Many of the optimizations for spherical harmonics rely on the
symmetry of the spherical harmonic functions.  Future work should
consider alternate quadrature sets that directly exploit real-space symmetry
to reduce the scaling of these methods with expansion order.
Working with distribution functions on the sphere also presents a new approach to
tensor analysis, traditional boundary value problems and extensions of black-box
multipole methods to potentials governed by more complicated PDEs.

  Three novel results that come from this connection are
space-optimal, vector-based representations for Cartesian polytensors,
an exact representation of point multipoles using discrete point charges,
and simplification of the moment shifting formulas for FMM.


%

\section*{Acknowledgments}
  Support for this work was provided by the USF Foundation.
  Line integral convolution routines used here were developed by Anne Archibald based on the work of Cabral, Brian and Laeith Leedom (SIGGRAPH '93: 263-270, 1993), and wrapped in scipy\cite{scipy} by David Huard.

\appendix 

\section{ Comparison to Spherical Harmonic Expressions}\label{s:harm}

  The expressions given in Sec.~\ref{s:fmm} are real-space counterparts of
the outer and inner expansions
obtained in the spherical harmonic formalism (Eq.~\ref{e:corresp}),\cite{cwhit94}
\begin{align}
\omega_{lm}(Q;A) &\equiv \int \rho(y)O_{lm}(y) \; d^3y \notag \\
  & = \int_S O_{lm}(\hat r) \sigma_o(\hat r) \; d^2r \label{e:trans} \\
\sigma_o(\hat r) &= \sum_{l=0}^{p-1}\sum_{m=-l}^l \frac{2l+1}{4\pi} M_{lm}(\hat r) \omega_{lm}(Q;A) \\
\mu_{lm}(Q;A) &\equiv \int \rho(y) M_{lm}(y) \; d^3y \notag \\
  & = \int_S M_{lm}(\hat r) \sigma_i(\hat r) \; d^2r \\
\sigma_i(\hat r) &= \sum_{l=0}^{p-1}\sum_{m=-l}^l \frac{2l+1}{4\pi} O_{lm}(\hat r) \mu_{lm}(Q;A)
.
\end{align}
As explained in Sec.~\ref{s:intro}, the integrals on $S$ can be carried out with exact summations over any
$O(p^2)$ points using quadrature methods capable of integrating all polynomials on $S$ up to order $2p-2$.
Both the number of degrees of freedom and the set of moments that can be represented
match between quadrature and spherical harmonic methods.  In addition, the na\"{\i}ve implementation of both methods
has the same scaling.  However, where harmonics require computation of $p^2$
basis functions, $O_{nm}$, the quadrature method only requires computation of the $p$ Legendre
polynomials, $L_n$.

  The shift formula of Eq.~\ref{e:shift} coincides with Ref.~\cite{cwhit94}, as can be seen
from the translations in Eq.~\ref{e:trans}
\begin{widetext}
\begin{align}
\sigma_o(R_1\hat r; x_1) &= \sum_{lm} \frac{2l+1}{4\pi} M_{lm}(\hat r) \sum_{jk}^l A^{lm}_{jk}((x_0-x_1)/R_1)
  \int_S O_{jk}(\tfrac{R_0}{R_1}\hat s) \sigma_o(R_0\hat s; x_0) \; d^2\hat s \\
 &= \int_S K(\tfrac{R_0\hat s + x_0-x_1}{R_1}, \hat r) \sigma_o(R_0\hat s; x_0) \; d^2\hat s , \\
 \intertext{since $A^{lm}_{jk}$ was defined so that}
 & \sum_{jk}^l A^{lm}_{jk}(t) O_{jk}(a) = O_{lm}(a+t) .
\end{align}
\end{widetext}

  To derive the other shift operators, define $B$ and $C$ as in Ref.~\cite{cwhit94} so that
\begin{widetext}
\begin{align}
M_{lm}(a-t) &= \sum^\infty_{jk} B^{lm}_{jk}(t) O_{jk}(a) \label{e:OIsum} \\
M_{lm}(a-t) &= \sum^\infty_{jk} C^{lm}_{jk}(t) M_{jk}(a), \label{e:IIsum} \\
\intertext{and note that}
\sigma_i(R_1 \hat r; x_0+t) &= \sum_{lm} \frac{2l+1}{4\pi} O_{lm}(\hat r) \int \sum^\infty_{jk} B^{lm}_{jk}(t/R_1) O_{jk}(\tfrac{R_0}{R_1}\hat s) \sigma_o(R_0 \hat s; x_0) \; d^2\hat s \\
 &= \int K(\hat r, \tfrac{R_0\hat s - t}{R_1}) \sigma_o(R_0\hat s; x_0) \; d^2\hat s \\
\sigma_i(R_1 \hat r; x_0+t) &= \sum_{lm} \frac{2l+1}{4\pi} O_{lm}(\hat r) \int \sum^\infty_{jk} C^{lm}_{jk}(t/R_1) M_{jk}(\tfrac{R_0}{R_1}\hat s) \sigma_i(R_0 \hat s; x_0) \; d^2\hat s \\
 &= \int K(\hat r, \tfrac{R_0\hat s - t}{R_1}) \sigma_i(R_0 \hat s; x_0) \; d^2\hat s
 .
\end{align}
\end{widetext}

\section{ Cartesian Tensors}\label{s:tens}

  This section derives several relations showing the identity between supersymmetric
tensors of order $n$ and homogeneous polynomials of the same order.  These prove
that relations involving such tensors can be stated identically in terms of polynomial functions.
Tensors are traditionally introduced via the Taylor expansion,
\begin{align}
G(r) &= \sum_{n=0}^\infty \frac{1}{n!} (r-r_0)^{(n)} \tdot{n} \partial_r^{(n)} G(r)\Big |_{r=r_0} \label{e:taylor} \\
  &= \sum_{n=0}^\infty \frac{1}{n!} ((r-r_0)\cdot \partial_s)^{n} G(s)\Big |_{s=r_0} \notag
,
\end{align}
where variables with parenthesized superscripts denote tensors.
Vectors are always written in lower-case.
The $n$-fold tensor contraction between $A^{(a)}$ and $B^{(b)}$
is denoted as $A^{(a)} \tdot{n} B^{(b)}$.

\subsection{ Translations between polynomials and symmetric tensors}

  We define the $n^\text{th}$ order directional moment of a three-dimensional
distribution, $\rho$, centered at the origin as the quantity,
\begin{equation}
m_n(r) = \int (r\cdot x)^n \rho(x) \; |d^3x| \label{e:mom}
\end{equation}
The distribution need not be positive, so $\rho |d^3x|$ is a signed measure.
It is clear from this definition that $m_n(r)$ is a homogeneous polynomial
of order $n$, with the monomial expansion,
\begin{align*}
m_n(r) = \sum_{n_1+n_2+n_3=n} \binom{n}{n_1 n_2 n_3} & r_x^{n_1} r_y^{n_2} r_z^{n_3} \\
     & \times M[n_1, n_2, n_3]
,
\end{align*}
where $M[n_1,n_2,n_3]$ are the monomial moments of $\rho$.
Since the test monomials, $x^{n_1}y^{n_2}z^{n_3}$ form a linearly
independent set, $m_n(r)$ spans the complete set of $\binom{n+2}{n}$ homogeneous polynomials in $r$.
These moments are identical to Applequist's reduced notation for the
supersymmetric tensor, $M^{(n)} = \partial_r^{(n)} m_n(r)/n! = \int x^{(n)} \rho(x) \; |d^3x|$.\cite{jappl89}  Alternatively,
to every order $n$ supersymmetric tensor, $A^{(n)}$, there corresponds
an order $n$ homogeneous polynomial,
\begin{equation}
a_n(r) = r^{(n)} \tdot{n} A^{(n)}
.
\end{equation}

  The symmetric tensor (outer) product can be found very easily using these polynomials as,
\begin{equation}
r^{(n+m)} \tdot{n+m} \mathcal S [ A^{(n)} \otimes B^{(m)} ] = a_n(r) b_m(r)
,
\end{equation}
where the symmetrizing operator, $\mathcal S$, averages over all $(n+m)!$ permutations
of the $n+m$ tensor indices.
This formula can be proven by noting that
\begin{align*}
r^{(n+m)} \tdot{n+m} &\left( A^{(n)} \otimes B^{(m)} \right) \\
  &= (r^{(n)} \tdot{n} A^{(n)}) (r^{(m)}\tdot{m} B^{(m)})
\end{align*}
for any permutation of the indices, since $r^{(n+m)}$ is symmetric.
The tensor contraction (inner product) is,\cite{cdebo92,mgasc00}
\begin{equation}
A^{(n)} \tdot{n} B^{(n)} = a_n(\partial_r) b_n(r) / n!
.
\end{equation}
These form a complete set of operations, since supersymmetric tensors can
be built from multiplying polynomials (forming symmetric tensor products),
and partial contractions can be found {\em via} differentiation, as
\begin{align*}
r^{(m)} \tdot{m}  &\left( A^{(n)} \tdot{n} B^{(n+m)} \right) \\
 &= (A^{(n)}\otimes r^{(m)}) \tdot{n+m} B^{(n+m)} \\
 &= \frac{1}{(n+m)!} (r\cdot\partial_s)^m a_n(\partial_s) b_{n+m}(s) \\
 &= \frac{1}{(n+m)!}  r^{(m)} \tdot{m} \partial_s^{(m)} a_n(\partial_s) b_{n+m}(s) \\
 &= \frac{m!}{(n+m)!} a_n(\partial_r) b_{n+m}(r) 
.
\end{align*}

  These identities are even simpler to manipulate when expressed in terms of
moment integrals,
\begin{align*}
&r^{(m)} \tdot{m}  \left( A^{(n)} \tdot{n} B^{(n+m)} \right) \\
 &= \iint (x\cdot y)^n (r\cdot y)^m \rho_A(x) \rho_B(y) \; |d^3x| |d^3y| \\
 &= \frac{m!}{(n+m)!} \iint (x\cdot \partial_r)^n  (r\cdot y)^{n+m} \rho_A(x) \rho_B(y) \; |d^3x| |d^3y|
 .
\end{align*}
This expansion always valid, since a signed measure yielding any tensor $M[n_1,n_2,n_3]$
(by Eq.~\ref{e:mom}) always exists due to the linear independence of the monomials mentioned above.

  Supersymmetric cartesian tensors and polynomial functions where each term has the same degree (termed projective polynomials), are thus in 1:1 correspondence.  A suitable quadrature formula for polynomials could reduce integrals such as the above to summation over point sets.
  
    In this work, we restrict our attention to polynomial functions defined over the surface of a sphere.
Because the Green's function, $1/r$, maintains the same functional form on scaling
the distance from the origin, $r$, only a subset of the full moment space is required for
describing solutions to Poisson's equation from sources at the origin.
That subspace is equivalent to the set of unique polynomials on a sphere of any radius.
Using this constraint ($x^2+y^2+z^2=R^2$) reduces the $\binom{n+2}{n}$ homogeneous polynomials of degree $n$ to only $2n+1$ linearly independent polynomials.

\subsection{ Representation of Polytensor Space}

  Analogous to Dette and Studden\cite{hdett97}, define the $p$-tuple of moments
for a given distribution as
\begin{align}
\mathbf c_p(\sigma) &= (m_0(r;\sigma), m_1(r;\sigma), \ldots, m_{p-1}(r;\sigma)) \label{e:cp}
\intertext{where}
&m_n(r;\sigma) \equiv \int_S (r\cdot \hat x)^n \sigma(x) d^2\hat x, \notag
\end{align}
are the moments of the distribution $\sigma$, integrated
over $S$, the surface of the unit sphere.
Since each individual moment, $m_n$ is a polynomial in $r$ corresponding to a tensor
of order $n$, $\mathbf c_p$ corresponds to a polytensor.\cite{jappl89}
The set of moments generated by a point mass at $\hat r_i$ is,
\begin{equation}
\mathbf c_p(\hat r_i) \equiv ((r\cdot \hat r_i)^0, (r\cdot \hat r_i)^1, \ldots, (r\cdot \hat r_i)^{p-1})
.
\end{equation}

  The $p$-th spherical moment space is given by, $\mathbf c_p(\sigma)$,
and represents the set of all polytensors up to order $p-1$ that can be expressed
as integrals of $r^{(n)}$ over a signed measure on the sphere (as in Eq.~\ref{e:cp}).
The space is convex and of dimension $p^2$.
For positive measures, it can be investigated using the techniques of Ref.~\cite{hdett97}.
For signed measures, it is isomorphic to the $p^2$ dimensional vector space, $\mathbb R^{p^2}$.
Further, using the reproducing kernel (Eq.~\ref{e:K}), it is easy to see that every spherical
polytensor $c_p$ admits a representation as a weighted sum over polytensors
generated by fixed unit vectors,
\begin{equation}
\mathbf c_p = \sum_{i=1}^N q_i \mathbf c_p(\hat r_i) \label{e:crep}
\end{equation}
where $N \le p^2$.

  This representation theorem shows that rather than the spherical harmonic coefficients
usually used to describe moments, we can use a set of $O(p^2)$ weights,
$w^\sigma_i = \sigma(\hat r_i) w^0_i$ associated to unit vectors, $\{\hat r_i\}$ in real space
to represent Cartesian polytensors up to order $p-1$.  Every operation on the Cartesian
tensor can be translated to an equivalent operation using the polynomial representation of Eq.~\ref{e:crep}.
In particular, any $n$-th order trace-free multipole tensor can be expressed in
terms of only $\sim \frac{3}{2} p^2$ quadrature points,
\begin{equation}
M^{(n)} = \sum_{i=1}^N q_i \hat r_i^{(n)} \label{e:Mrep}
.
\end{equation}
The natural connection between polynomials and tensors
developed in the previous section simplifies the use of this set for function expansions.

  In summary, there is a direct correspondence between supersymmetric tensors
and polynomials.  This connection is usually used to generate reduced representations
of supersymmetric tensors in terms of coefficients, $M[n_1,n_2,n_3]$,
on basis monomials, $r_x^{n_1} r_y^{n_2} r_z^{n_3}$.\cite{jappl89,gdass02,abord03,pren03}
This representation is cumbersome because it requires a particular choice for the orientation
of the coordinate axes, and (for scale-free problems) because it is not
specialized for trace-tree polynomials.  Using Eq.~\ref{e:rep} and Eq.~\ref{e:Mrep},
quadrature rules for integrating polynomial functions give the desired
compact and geometrically useful representation in terms of values taken by
$m_n(r_i)$ or $\rho(r_i)$ at a small set of points, $r_i$, in real space.
The basis vectors, $\{\hat r_i\}$, are much more computationally useful
because they can be individually rotated / translated
and their coefficients can be directly multiplied
without re-factoring between monomials or computing Clebsch-Gordon coefficients.

\bibliography{lit}

\end{document}